\documentclass[square]{ws-procs11x85}
\usepackage{balance} 
\usepackage{graphicx}

\def\Journal#1#2#3#4{{#1} {\bf #2}, #3 (#4)}

\newcommand{\met}{\hbox{E\kern-0.5em\lower-0.1ex\hbox{/}}_T}

\begin{document}

\twocolumn[
\title{Optical depth for VHE $\gamma$-rays from distant sources from a generic EBL density}

\author{M. Raue}

\address{Max-Planck-Institut f\"ur Kernphysik, Heidelberg, Germany}

\author{D. Mazin}

\address{IFAE, Edifici Cn. Universitat Autonoma de Barcelona, Barcelona, Spain}


\begin{abstract}
Very-high-energy (VHE; $E>100$\,GeV) $\gamma$-rays from distant sources suffer attenuation through pair-production with low energy photons from the diffuse extragalactic photon fields in the ultraviolet (UV) to far-infrared (FIR) (commonly referred to as Extragalactic Background Light; EBL). When modeling the intrinsic spectra of the VHE $\gamma$-ray sources it is crucial to correctly account for the attenuation. Unfortunately, direct measurements of the EBL are difficult and the knowledge about the EBL over certain wavelength ranges is poor. To calculate the EBL attenuation usually predictions from theoretical models are used. Recently, the limits on the EBL from direct and indirect methods have narrowed down the possible EBL range and many of the previous models are in conflict with these limits. We propose a new generic EBL density (not a complete model), which is in compliance with the new EBL limits. EBL evolution with redshift is included in the calculation in a very simple but effective ad-hoc way. Properties of this generic EBL are discussed.
\end{abstract}
\vskip12pt  
]

\bodymatter

\section{Introduction}

The EBL consists of all photons emitted by stars, partially reprocessed by dust, integrated over the history of the universe. VHE $\gamma$-rays from distant sources interact with this photon field via pair-production and the flux is attenuated \cite{nikishov:1962a}. The pair-production cross-section is strongly peaked, resulting in an energy dependent attenuation signature in the measured VHE $\gamma$-ray spectra. When modeling intrinsic source spectra, this attenuation effect has to be taken into account. If the optical depth $\tau (E, z)$ from the EBL attenuation is known for an energy $E$ and redshift $z$, the intrinsic spectrum $F_{\mathrm{int}}(E)$ can be calculated: $F_{\mathrm{int}}(E) = F_{\mathrm{obs}}(E)\, e^{\tau(E, z)}$.

Directly measuring todays EBL, especially in the mid-infrared (MIR) wavelength region, is difficult due to dominant foregrounds. Strict lower limits are derived from source counts and rather loose upper limits come from direct measurements (see \cite{hauser:2001a} for a review). An indirect method of determining the EBL utilizes the VHE $\gamma$-ray spectra from distant sources. With basic assumptions about the source physics (and thereby about the intrinsic spectrum), limits on the EBL can be derived \cite{stecker:1992a}.

Recently, the limits on the EBL have tightened: source counts from the SPITZER instrument gave stringent lower limits in the previously largely unexplored near-infrared (NIR; 2-10\,$\mu$m) wavelength region \cite{papovich:2004a,fazio:2004a,frayer:2006a}. The detection of several distant VHE $\gamma$-ray sources with the HESS experiment led to strong upper limits on the EBL in the same wavelength region\cite{aharonian:2006:hess:ebl:nature,aharonian:2007:hess:0229}. \cite{mazin:2007a} re-examined the spectra of all known VHE blazars and derived limits on the EBL over a wide wavelength range.

In this article we aim to construct an EBL density (not a complete model), which is in compliance with all the latest EBL limits. For source at redshifts $z \sim 0.2$ the evolution of the EBL has already an measurable effect on the attenuation. Here we follow a very simple but effective ad-hoc approach to include the EBL evolution into the calculation.

\section{A new generic EBL}
 
\begin{figure*}[tb]
\centering{\includegraphics[width=\textwidth]{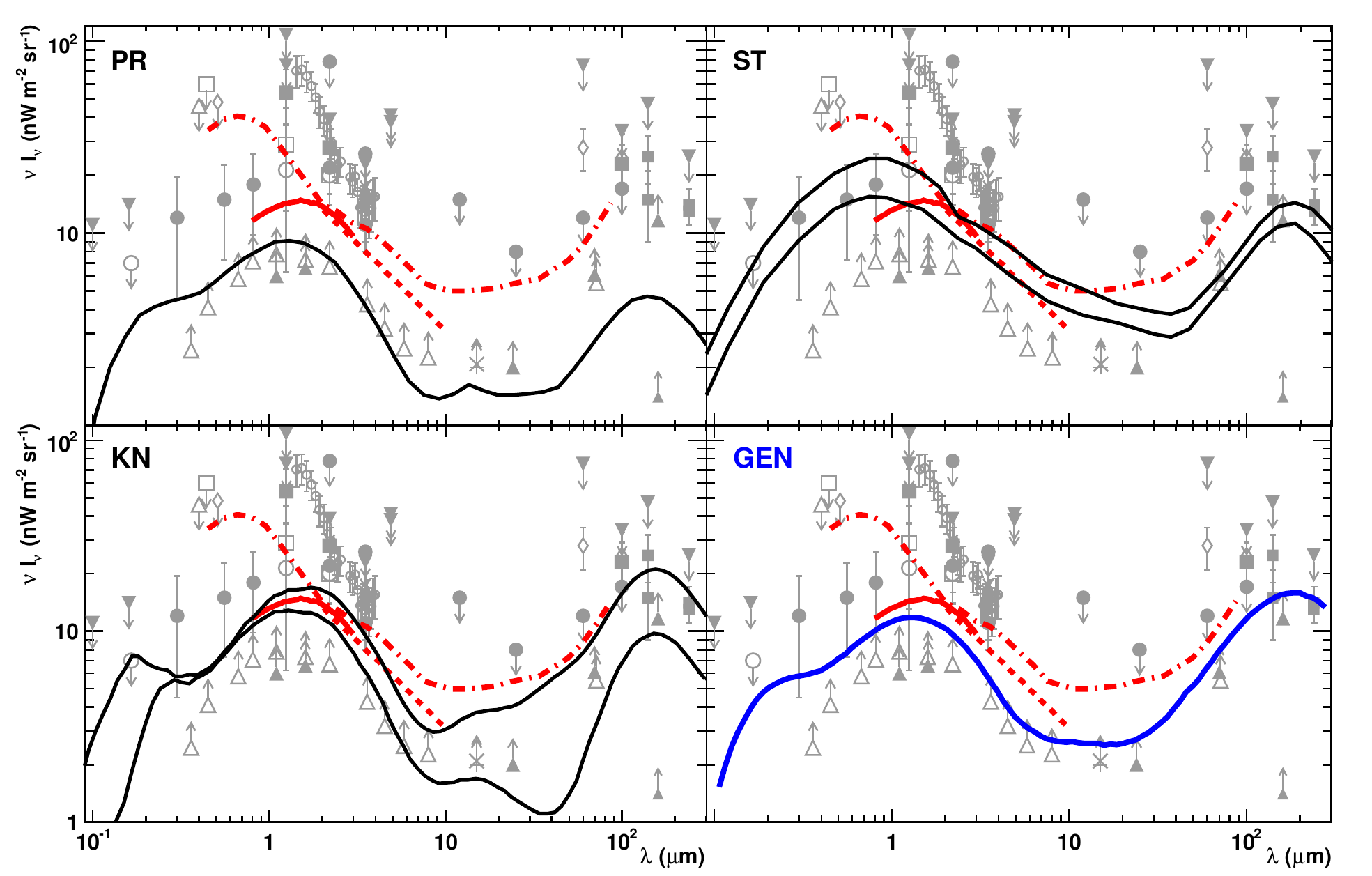}}
\caption{\label{fig:ebl_lim_vs_ebl_models} EBL models versus EBL limits. Black solid curves show different EBL models for $z=0$. Grey markers are measurements and limits from direct measurements, fluctuation analysis and from source counts (data compilation from \cite{mazin:2007a}). Red curves are upper limits derived from VHE $\gamma$-ray spectra (solid\cite{aharonian:2006:hess:ebl:nature}, dashed\cite{aharonian:2007:hess:0229}, dashed-dotted\cite{mazin:2007a}). \textit{Upper-Right:} EBL model from \cite{primack:2005a}. \textit{Upper-Left:} Fast evolution and baseline models from \cite{stecker:2006a}. \textit{Lower-Left:} Updated high and low models from \cite{kneiske:2002a}. \textit{Lower-Right:} Generic EBL proposed in this paper (blue curve).}
\end{figure*}

A large number of different EBL models exists. In the following, we will discuss three widely used EBL models from \cite{primack:2005a} (PR), \cite{malkan:2001a,stecker:2006a} (ST) and  \cite{kneiske:2002a} (KN) in the light of the new limits on the EBL (for other models see e.g. \cite{franceschini:2001a,totani:2002a}). All three models have been developed before the recent limits on the EBL from VHE $\gamma$-ray spectra and SPITZER source counts have been published. In Fig.~\ref{fig:ebl_lim_vs_ebl_models} predictions for the EBL at redshift $z = 0$ calculated with these EBL models are shown in comparison to the EBL limits (details are given in the caption).

The PR model (Fig.~\ref{fig:ebl_lim_vs_ebl_models} upper left panel) is well below the upper limits, but falls below the lower limits from source counts in the NIR. The ST fast evolution model (Fig.~\ref{fig:ebl_lim_vs_ebl_models} upper right panel) is marginally above the upper limits, while the baseline model is below them. Both models have a hard spectral slope in the NIR-MIR, which is excluded by \cite{aharonian:2007:hess:0229} using the hard VHE $\gamma$-ray spectrum from 1ES\,0229+200 (but see also \cite{stecker:2007a,stecker:2008a} for an alternative interpretation). The KN high model (Fig.~\ref{fig:ebl_lim_vs_ebl_models} lower left panel) is in conflict with the upper limits while the low model falls below the lower limits in the NIR-MIR (recently, a new version of the model has been published \cite{kneiske:2007a}). We note that in general there is a rather large spread in the prediction of the EBL from different models, reflecting the uncertainties in the knowledge of the different parameters and components contributing to the EBL (especially in the MIR to FIR).

Given these uncertainties and the conflicts between models and limits we propose to utilize a new generic EBL density for VHE $\gamma$-ray attenuation calculations. Other generic EBL densities have already been proposed in e.g. \cite{aharonian:2002a} and \cite{dwek:2005a}, to address similar problems. The EBL density is constructed to be in compliance with the current EBL limits and lying just above the lower limits from source counts. In the optical to NIR there is not much room left between upper and lower limits and we adopt a scaled PR model ($z = 0$). In the FIR we use a scaled KN model and connect the two shapes smoothly in the MIR, with the EBL density on the level of the source counts lower limits. The resulting EBL density is shown in the lower right panel of Fig.~\ref{fig:ebl_lim_vs_ebl_models}.

\section{Generic EBL evolution}

\begin{figure*}[tb]
\centering{\includegraphics[width=\textwidth]{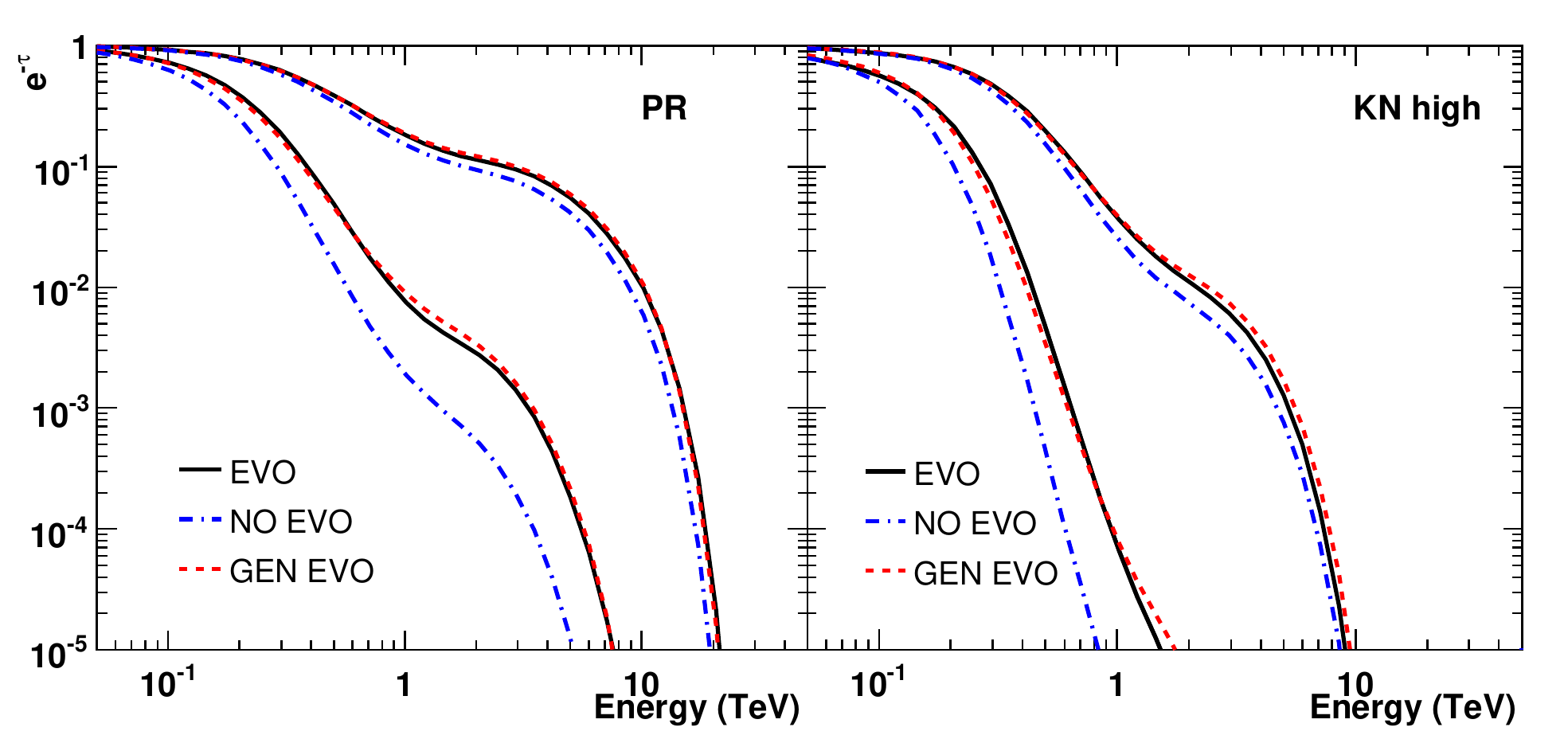}}
\caption{\label{fig:modelEvo_vs_genEvo_att} Attenuation $e ^{-\tau}$ for VHE $\gamma$-rays emitted at redshift $z = 0.2$ and $z = 0.5$ (higher and lower curves respectively) calculated using (a) the original EBL model including EBL evolution (solid black; EVO), (b) the EBL model without EBL evolution (dashed-dotted blue; NO EVO) and (c) same as (b) but including generic EBL evolution with $f_{\mathrm{evo}} = 1.2$ (dashed red; GEN EVO; partially overlapping with (a)). \textit{Left:} EBL model from PR. \textit{Right:} High model from KN.}
\end{figure*}
 
\begin{figure}[htb]
\centering{\includegraphics[width=0.5\textwidth]{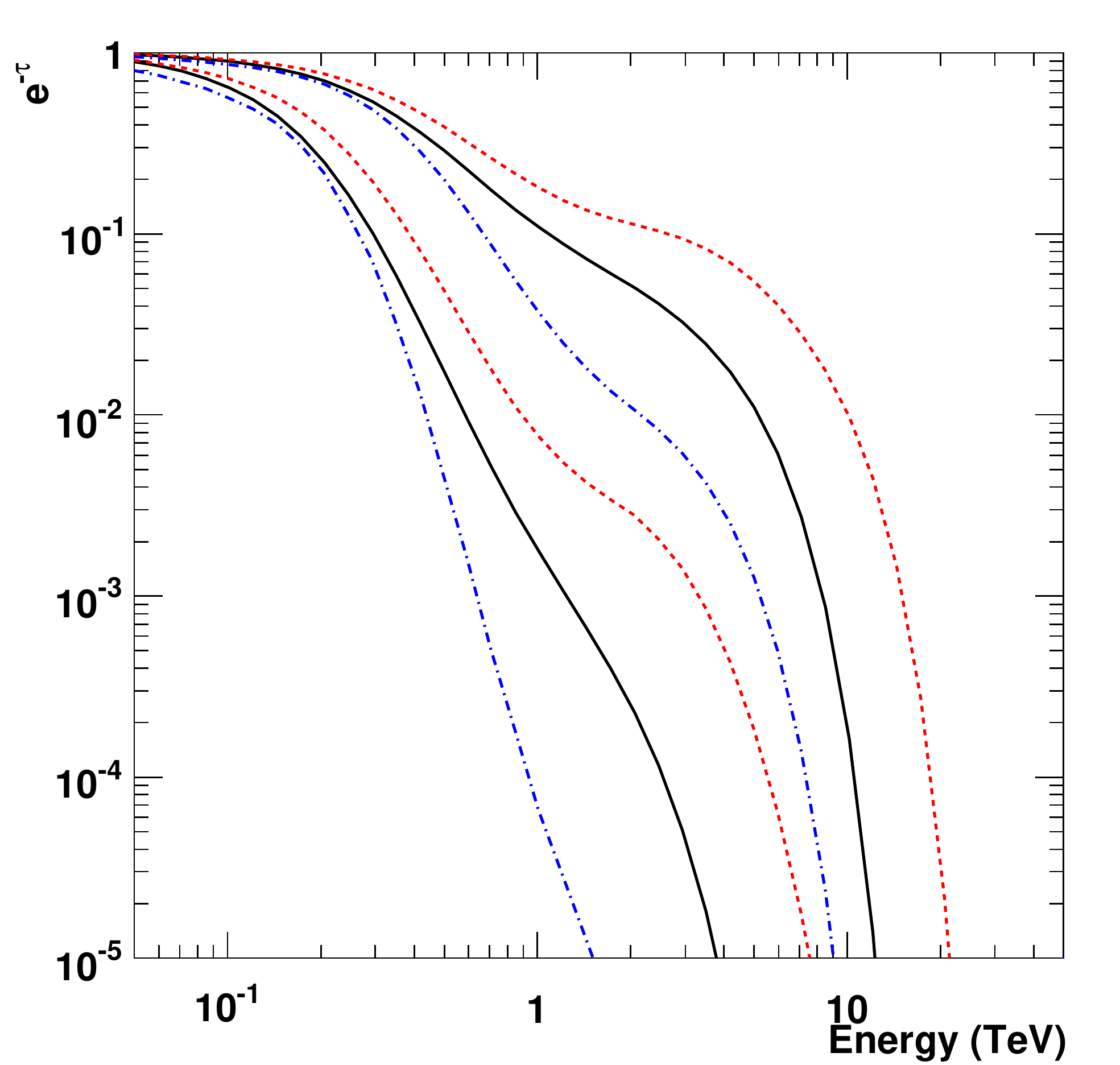}}
\caption{\label{fig:eblAtt_GEN01} Attenuation $e^{-\tau}$ for VHE $\gamma$-rays emitted at redshift $z = 0.2$ and $z = 0.5$  (higher and lower curves respectively) calculated for the generic EBL with  $f_{\mathrm{evo}} = 1.2$ (black solid curves) in comparison to the PR EBL model (red dashed curves) and the high model from KN (blue dashed-dotted curves).}
\end{figure}

The EBL is not produced instantaneously at a distant redshift but builds up slowly over the history of the universe (EBL evolution), tracing the star formation history. The redshift integrated photon number density is therefore lower than in the static case, resulting in a higher attenuation in case the EBL evolution is not taken into account. In Fig.~\ref{fig:modelEvo_vs_genEvo_att} the attenuation ($e ^{-\tau}$ ) resulting from an EBL model including EBL evolution is compared to the attenuation calculated for the same model without EBL evolution. The optical depth $\tau$ in the energy range from 50\,GeV to 50\,TeV resulting from the model without EBL evolution is between 10-15\% ($z = 0.2$) and 20-35\% ($z = 0.5$) higher than for the model with EBL evolution.\footnote{The offset is given in $\tau$, since the rather constant relative offset in $\tau$ transfers in a wide range of attenuation $e^{-\tau}$ depending on the actual value of $\tau$. Since for modeling the relevant quantity is the attenuation, we show $e^{-\tau}$ in the figures.}

The generic EBL discussed in the previous section is defined for $z = 0$ and does not have a known redshift evolution. A simple first order approach to include EBL evolution in the calculation is to change the cosmological photon number density scaling  from $n(\epsilon,z) \propto (1 + z)^{3}$ to $(1 + z)^{3 - f_{\mathrm{evo}}}$ adding $f_{\mathrm{evo}} > 0$ (e.g. \cite{madau:1996a} and recently revived by \cite{aharonian:2007:hess:0229}). The factor $f_{\mathrm{evo}}$ can be tuned to match the EBL evolution (as derived from complete EBL models) for different redshift ranges. Here we choose $f_{\mathrm{evo}} = 1.2$, which results in a good match for redshift up to $z \sim 0.7$.  In Fig.~\ref{fig:modelEvo_vs_genEvo_att} the attenuation calculated from the EBL models (including EBL evolution) is compared to the attenuation derived using the generic EBL density and generic EBL evolution with $f_{\mathrm{evo}} = 1.2$. The agreement is good, resulting in a deviation in $\tau$ of $< 4$\% for $z = 0.2$ and $< 10$\% for $z = 0.5$.

The attenuation resulting from the generic EBL with $f_{\mathrm{evo}} = 1.2$ in comparison to the attenuation from the PR and the KN high model is shown in Fig.~\ref{fig:eblAtt_GEN01}. As the generic EBL density is situated between the two EBL models,  the attenuation resulting from the generic EBL lies between the two models as well. The difference in tau between the generic EBL density and the PR and KN EBL models in the energy range 100\,GeV to 10\,TeV is in the order of 20\% to 50\%.

\section{Conclusions}

The proposed generic EBL density together with generic EBL evolution provides an up-to-date tool for source modeling of distant VHE $\gamma$-ray sources, taking into account the recent EBL limits. Attenuation tables for different redshifts will be supplied online (http://www.desy.de/$\sim$mraue/ebl/). We want to stress that the generic EBL is not a complete theoretical model but a simple fit to the current limits.

Given that there is still some spread in the possible EBL realizations, certain uses might require an absolute lower and upper limit EBL density. Here, scaled version of the generic EBL density could provide reasonable results ($n'(\epsilon) = f \cdot n(\epsilon)$, with e.g. $f = 0.8$ and $f = 1.3$), though there is some freedom of choice in the MIR to FIR.

The generic EBL evolution can be applied to any EBL density, e.g. to reduce the systematic error in EBL studies like \cite{mazin:2007a}.
\section*{Acknowledgments}
{\small
This research has made use of NASA's Astrophysics Data System.
}
\balance



\def\Journal#1#2#3#4{{#4}, {#1}, {#2}, #3}
\def\NAT{Nature}
\def\AAA{A\&A}
\def\ApJ{ApJ}
\def\AJ{Astronom. Journal}
\def\Aph{Astropart. Phys.}
\def\ApJS{ApJSS}
\def\MNRAS{MNRAS}
\def\NIM{Nucl. Instrum. Methods}
\def\NIMA{Nucl. Instrum. Methods A}
\def\aj{AJ}%
\def\actaa{Acta Astron.}%
\def\araa{ARA\&A}%
\def\apj{ApJ}%
\def\apjl{ApJ}%
\def\apjs{ApJS}%
\def\ao{Appl.~Opt.}%
\def\apss{Ap\&SS}%
\def\aap{A\&A}%
\def\aapr{A\&A~Rev.}%
\def\aaps{A\&AS}%
\def\azh{AZh}%
\def\baas{BAAS}%
\def\bac{Bull. astr. Inst. Czechosl.}%
\def\caa{Chinese Astron. Astrophys.}%
\def\cjaa{Chinese J. Astron. Astrophys.}%
\def\icarus{Icarus}%
\def\jcap{J. Cosmology Astropart. Phys.}%
\def\jrasc{JRASC}%
\def\mnras{MNRAS}%
\def\memras{MmRAS}%
\def\na{New A}%
\def\nar{New A Rev.}%
\def\pasa{PASA}%
\def\pra{Phys.~Rev.~A}%
\def\prb{Phys.~Rev.~B}%
\def\prc{Phys.~Rev.~C}%
\def\prd{Phys.~Rev.~D}%
\def\pre{Phys.~Rev.~E}%
\def\prl{Phys.~Rev.~Lett.}%
\def\pasp{PASP}%
\def\pasj{PASJ}%
\def\qjras{QJRAS}%
\def\rmxaa{Rev. Mexicana Astron. Astrofis.}%
\def\skytel{S\&T}%
\def\solphys{Sol.~Phys.}%
\def\sovast{Soviet~Ast.}%
\def\ssr{Space~Sci.~Rev.}%
\def\zap{ZAp}%
\def\nat{Nature}%
\def\iaucirc{IAU~Circ.}%
\def\aplett{Astrophys.~Lett.}%
\def\apspr{Astrophys.~Space~Phys.~Res.}%
\def\bain{Bull.~Astron.~Inst.~Netherlands}%
\def\fcp{Fund.~Cosmic~Phys.}%
\def\gca{Geochim.~Cosmochim.~Acta}%
\def\grl{Geophys.~Res.~Lett.}%
\def\jcp{J.~Chem.~Phys.}%
\def\jgr{J.~Geophys.~Res.}%
\def\jqsrt{J.~Quant.~Spec.~Radiat.~Transf.}%
\def\memsai{Mem.~Soc.~Astron.~Italiana}%
\def\nphysa{Nucl.~Phys.~A}%
\def\physrep{Phys.~Rep.}%
\def\physscr{Phys.~Scr}%
\def\planss{Planet.~Space~Sci.}%
\def\procspie{Proc.~SPIE}%
          


\end{document}